\documentclass[aps,prd,preprint,superscriptaddress,amsmath,amssymb,showpacs]{revtex4-1}
\usepackage{dcolumn}
\usepackage{graphicx}
\usepackage{float}
\usepackage{physics}
\usepackage[colorlinks=true,allcolors=blue]{hyperref}
\usepackage{mathrsfs}
\usepackage{inputenc}
\usepackage{xcolor}

\newcommand{\be}{\begin{equation}}
\newcommand{\ee}{\end{equation}}
\newcommand{\bea}{\begin{eqnarray}}
\newcommand{\eea}{\end{eqnarray}}

\newcommand{\bln}{\begin{align}}
\newcommand{\eln}{\end{align}}
\newcommand{\bst}{\begin{split}}
\newcommand{\est}{\end{split}}
\newcommand{\bi}{\begin{itemize}}
\newcommand{\ei}{\end{itemize}}
\newcommand{\bn}{\begin{enumerate}}
\newcommand{\en}{\end{enumerate}}

\def\lad{L} 
\def\lg{\lam_{GB}}

\def\ov{\over}
\def\le{\left}
\def\ri{\right}

\def\lam{{\lambda}}
\def\Lam{{\Lambda}}
\def\al{{\alpha}}

\def \lam {\lambda}

\def\apr{{\alpha'}}

\def\GeV{{\rm GeV}}

\def\lam{{\lambda}}

\def\eeq{\end{equation}}

\newcommand{\rev}[1]{#1}

\UseRawInputEncoding
\begin{document}
\title{Gauss-Bonnet corrections to Complexity Growth with a Probe String}

\author{Wen-Bin Chang}
\email{changwb@mails.ccnu.edu.cn}
\affiliation{College of Intelligent Systems Science and Engineering, Hubei Minzu University, Enshi 445000, People's Republic of China}

\date{\today}

\begin{abstract}
We investigate the effect of Gauss-Bonnet (GB) corrections on holographic complexity growth within the framework of the Complexity=Action (CA) conjecture. 
By introducing a probe string into a GB $AdS$ black brane background, we analyze the time derivative of the Nambu-Goto (NG) action as the holographic dual to complexity growth.
Our results indicate that the complexity growth is maximized for a stationary string and is suppressed by its motion. 
At fixed temperature, the stationary string complexity growth is independent of the GB coupling, whereas that of moving strings is suppressed by stronger GB corrections.
Finally, the growth rate is shown to increase linearly with temperature, confirming that higher temperatures systematically drive the complexity growth.

\end{abstract}

\maketitle

\section{Introduction}\label{sec:01_intro}
Recently, holographic formulations of both entanglement entropy and computational complexity have become essential for connecting quantum information concepts with the geometry of gravitational systems\cite{Watrous:2008any,Osborne_2012,PhysRevA.94.040302,Banerjee:2017qti,Takayanagi:2012kg,Hubeny:2007xt}.
In particular, the duality between the complexity of a boundary field theory and a gravitational quantity in the bulk, known as holographic complexity, provides an important framework for exploring the physics of black hole interiors\cite{Zhao:2017iul,Cazares:2012gxg,Hayden:2007cs,Susskind:2015toa}.
Several conjectures have been proposed for holographic complexity, among which the complexity=volume conjecture asserts that complexity is measured by the regularized volume of the maximal codimension-one hypersurface spanning the Einstein--Rosen bridge\cite{Stanford:2014jda}.
Another proposition is the complexity=action conjecture, which asserts that holographic complexity is equivalent to the action of a spacetime region known as the Wheeler-DeWitt (WDW) patch\cite{Brown:2015bva,Brown:2015lvg}.
The complexity=volume 2.0 conjecture identifies holographic complexity with the spacetime volume of the WDW patch, and has been shown to be consistent with the Lloyd bound in certain cases\cite{Couch:2016exn}.
Additionally, the complexity=anything conjecture posits that any member of an infinite class of gravitational observables, defined by integrating a scalar functional over an extremal codimension-one surface, serves as an equally viable candidate for the holographic dual of complexity by universally displaying its characteristic late-time linear growth and switchback effect\cite{Belin:2021bga,Belin:2022xmt,Jorstad:2023kmq,Myers:2024vve}.

As holographic complexity exhibits non-local behavior, its further investigation may rely on the deployment of non-local probes\cite{Moosa:2017yvt,Abad:2017cgl,Fu:2018kcp}.
By considering a Wilson line operator in $AdS$ spacetime associated with a fundamental string moving on a great circle in the submanifold, Nagasaki et al. investigated holographic complexity and systematically analyzed its dependence on parameters such as string velocity, charge, rotational effects, black hole mass, backreaction, and Horndeski interactions\cite{Nagasaki:2017kqe,Nagasaki:2018csh,Nagasaki:2019icm,Nagasaki:2022lll,Zhou:2021vsm,Santos:2020xox,Bravo-Gaete:2020lzs,Chang:2024muq,Guo:2025bgx}. 
Further progress on holographic complexity is presented in\cite{Guo:2017rul,Aguilar-Gutierrez:2023zqm,Pedraza:2021fgp,Chen:2023tpi,Aguilar-Gutierrez:2023ccv,Caceres:2024edr,Jiang:2025qai}.
Higher-derivative corrections arising from stringy or quantum effects, such as the GB term originating from the world-volume action of D7-branes, represent leading order $1/N_c$ corrections in string theory\cite{Douglas:2006es,Buchel:2008vz,Aharony:1999rz,Aharony:2007dj}.
Therefore, studying the effects of GB corrections on holographic complexity may provide valuable insights into the quantum aspects of gravity.
This work investigates the effect of GB corrections on the growth of holographic complexity in GB $AdS$ black brane background by introducing a probe string governed by the NG action. 
The time derivative of the action is interpreted as the complexity growth under the CA conjecture.

This paper is organized as follows. Sec.~\ref{sec:02} reviews the background geometry of GB gravity. In Sec.~\ref{sec:03}, we present the calculation of the complexity growth along with the numerical results and their analysis. Finally, Sec.~\ref{sec:04} summarizes our main conclusions.

\section{BACKGROUND GEOMETRY}\label{sec:02}

Einstein gravity can be extended by the GB term, incorporating a quadratic curvature correction that in string theory originates from the world-volume action of D7-branes\cite{Buchel:2008vz,Aharony:1999rz,Aharony:2007dj}.
The leading-order effective action for the gravity sector in $AdS_5$ can be written as\cite{Brigante:2007nu,Brigante:2008gz}
\begin{gather}
I= {1 \ov 16 \pi G_N} \int d^5 x \, \sqrt{- g} \le(R - 2 \Lambda +
\lad^2  \le(\al_1  R^2+  \al_2 R_{\mu \nu} R^{\mu \nu}+\al_3 R^{\mu
\nu\rho \sigma} R_{\mu \nu \rho\sigma} \ri)\ri) \ .
\end{gather}
Here, $\Lam = -{6 \ov \lad^2}$ and we take $\al_i \sim
{\apr \ov \lad^2} \ll 1$. 
$R$ denotes the Ricci scalar, $R_{\mu\nu}$ the Ricci tensor, and $R_{\mu\nu\rho\sigma}$ the Riemann tensor, with $G_N$ being the Newton constant.
GB gravity resolves the redefinition ambiguity of the couplings $\al_1$ and $\al_2$ by fixing all $\al_i$ in terms of a single parameter, $\lambda_{GB}$\cite{Brigante:2007nu,Brigante:2008gz}.
The action in GB gravity, inclusive of the $\lambda_{GB}$ term, is given by\cite{Brigante:2007nu,Zwiebach:1985uq}
\begin{equation}
\label{action} I = \frac{1}{16\pi G_N} \mathop\int{d^{5}x \,
\sqrt{-g} \, \le[R-2\Lambda+ {\lg \ov 2} \lad^2
(R^2-4R_{\mu\nu}R^{\mu\nu}+R_{\mu\nu\rho\sigma}R^{\mu\nu\rho\sigma})
\ri]} \ .
\end{equation}
A black brane solution exists in GB gravity and can be written as\cite{Cai:2001dz}
\begin{equation}
   ds^2=\frac{L^2}{z^2}(-af_{GB}(z)dt^2+d\vec x^2+\frac{1}{f_{GB}(z)}dz^2),
\end{equation}
where
\begin{equation}
   a=\frac{1}{2}(1+\sqrt{1-4\lambda_{GB}}),
\end{equation}
with
\begin{equation}
  f_{GB}(z)=\frac{1}{2\lambda_{GB}}(1-\sqrt{1-4\lambda_{GB}(1-\frac{z^4}{z_h^4})}).
\end{equation}
The boundary coordinates are given by $\vec{x} = x_1, x_2, x_3$ , $z$ corresponds to the fifth dimension, and $z_h$ marks the horizon.
In $(4+1)$-dimensional GB gravity, the causality condition requires $\lambda_{GB}\leq \frac{9}{100}$, as reported in\cite{Brigante:2008gz}.
The Hawking temperature of the black brane is given by\cite{Cai:2001dz}
\begin{equation}\label{eq13}
  T=\frac{\sqrt{a}}{\pi z_h}.
\end{equation}

\section{EVALUATION OF THE COMPLEXITY GROWTH}\label{sec:03}
Following the approach of \cite{Nagasaki:2017kqe}, we introduce a fundamental string into the bulk spacetime to realize a nonlocal Wilson line operator and analyze the growth of the NG action as the dual representation of complexity growth via the CA conjecture.
To investigate the motion of the string within a subspace of the given spacetime geometry, we parameterize its worldsheet by the coordinates $\tau$ and $\sigma$
\begin{equation}
\label{eqs}
\ t=\tau,\  \ r=\sigma, \ \phi=v\tau+\xi(\sigma).
\end{equation}
In this context, the function $\xi(\sigma)$ specifies the shape of the string and $v$ represents its velocity with respect to the black hole.
The time derivative of the NG action for the fundamental string is computed by integrating the square root of the determinant of the induced metric over the WDW patch
\begin{equation}
\label{eqt}
\frac{d S_{N G}}{d t}=T_{s} \int_{z_{h}}^{\infty} d z \sqrt{-g_{\text {ind }}(z)}=T_{s} \int_{z_{h}}^{\infty} d z \sqrt{\frac{L^4 \left(a f_{GB}(z) \left(\xi^{\prime 2} f_{GB}(z)+1\right)-v^2\right)}{f_{GB}(z) z^4}}.
\end{equation}
Here, $T_{s}$ denotes the tension of the fundamental string.
The equation of motion for $\xi$ is obtained by varying the above action
\begin{equation}
\label{equ}
\ \Pi_{\xi}=\frac{a\, \xi' f_{GB}(z)\, L^2}{z^2 \sqrt{a\, \xi^{\prime 2} f_{GB}(z) + a - \frac{v^2}{f_{GB}(z)}}},
\end{equation}
where $\Pi_{\xi}$ is the integration constant.
Solving the above equation yields
\begin{equation}
\label{eqv}
\xi'^{2}=\frac{\Pi^2_{\xi} z^4 \left(a f_{GB}(z) - v^2\right)}{a \left(f_{GB}(z)\right)^2 \left(a f_{GB}(z) L^4 - \Pi^2_{\xi} z^4\right)}.
\end{equation}
To ensure the expression yields real values, we require the numerator and denominator to change sign concurrently at the critical point, which is then determined as
\begin{equation}
\label{eqw}
a f_{GB}(z_c)=v^2,
\end{equation}
and
\begin{equation}
\label{eqx}
a f_{GB}(z_c) L^4 = \Pi^2_{\xi} z_c^4.
\end{equation}
By combining the results from (\ref{eqv}), (\ref{eqw}), and (\ref{eqx}) within (\ref{eqt}), the growth of the action is given by
\begin{equation}
\label{eqfn}
\frac{d S_{N G}}{d t}=T_s \int_{z_h}^{\infty} d z\sqrt{\frac{a L^4 z_c^4 \left(a f_{GB}(z) - v^2\right)}{z^4 \left(a f_{GB}(z) z_c^4 - v^2 z^4\right)}}.
\end{equation}

\begin{figure}[H]
    \centering
      \setlength{\abovecaptionskip}{-0.1cm}
    \includegraphics[width=7cm]{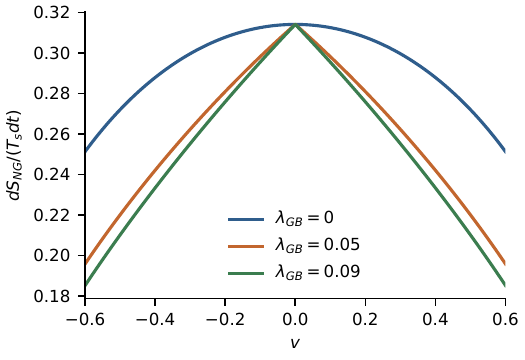}
    \caption{\label{fig1} \rev{Action growth versus string velocity for different $\lambda_{GB}$ at $T=0.1~\GeV$.}}
\end{figure}

In the following analysis, we numerically solve integral (\ref{eqfn}) to examine how GB corrections, string velocity, and temperature influence the growth of the NG action, as illustrated in the subsequent figures.
As shown in Fig.~\ref{fig1}, the dependence of action growth on string velocity was examined at fixed temperature for different values of $\lambda_{GB}$.
The action growth exhibits symmetry about $v=0$ for all values of $\lambda_{GB}$, according to Fig.~\ref{fig1}.
This observed symmetry indicates that the complexity growth depends solely on the magnitude of the string velocity, independent of its direction.
The action growth reaches its maximum when the string is stationary $(v=0)$ and decreases monotonically as the magnitude of the string velocity increases. 
This indicates that string motion tends to suppress the growth of complexity. 
Such behavior is consistent with previous findings, suggesting that it may be a universal property of holographic complexity\cite{Nagasaki:2018csh}.
\rev{For the static string, Eq.~(\ref{eqfn}) gives the analytic result}
\begin{equation}
\label{eqstatic}
\rev{\left.\frac{d S_{NG}}{d t}\right|_{v=0}=T_s L^2 \sqrt{a}\int_{z_h}^{\infty}\frac{dz}{z^2}=T_s L^2\frac{\sqrt{a}}{z_h}=T_s L^2 \pi T,}
\end{equation}
which is independent of $\lambda_{GB}$ at fixed temperature. 
This analytic result in Eq.~(\ref{eqstatic}) explains why all curves in Fig.~\ref{fig1} meet at the same value at $v = 0$.
For moving strings, increasing $\lambda_{GB}$ decreases the action growth over the velocity range, indicating that GB corrections suppress complexity growth.

\begin{figure}[H]
    \centering
      \setlength{\abovecaptionskip}{-0.1cm}
    \includegraphics[width=7cm]{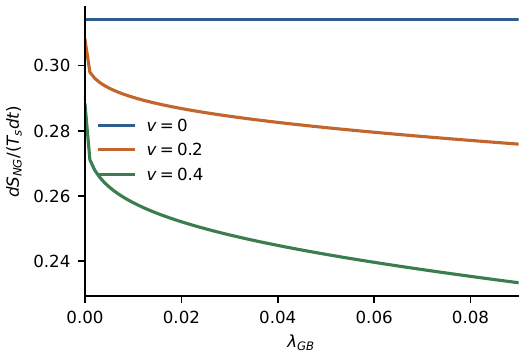}
    \caption{\label{fig2} Action growth versus $\lambda_{GB}$ for different string velocity at $T= 0.1GeV$.}
\end{figure}

To further elucidate the effect of the GB corrections on the complexity growth, we plot the NG action growth versus the GB coupling parameter $\lambda_{GB}$ for several fixed string velocities in Fig.~\ref{fig2}.
Figure~\ref{fig2} shows the same behavior from a complementary viewpoint. The $v=0$ result is independent of $\lambda_{GB}$, while the curves for $v=0.2$ and $v=0.4$ decrease as $\lambda_{GB}$ increases. 

\begin{figure}[H]
    \centering
      \setlength{\abovecaptionskip}{-0.1cm}
    \includegraphics[width=13cm]{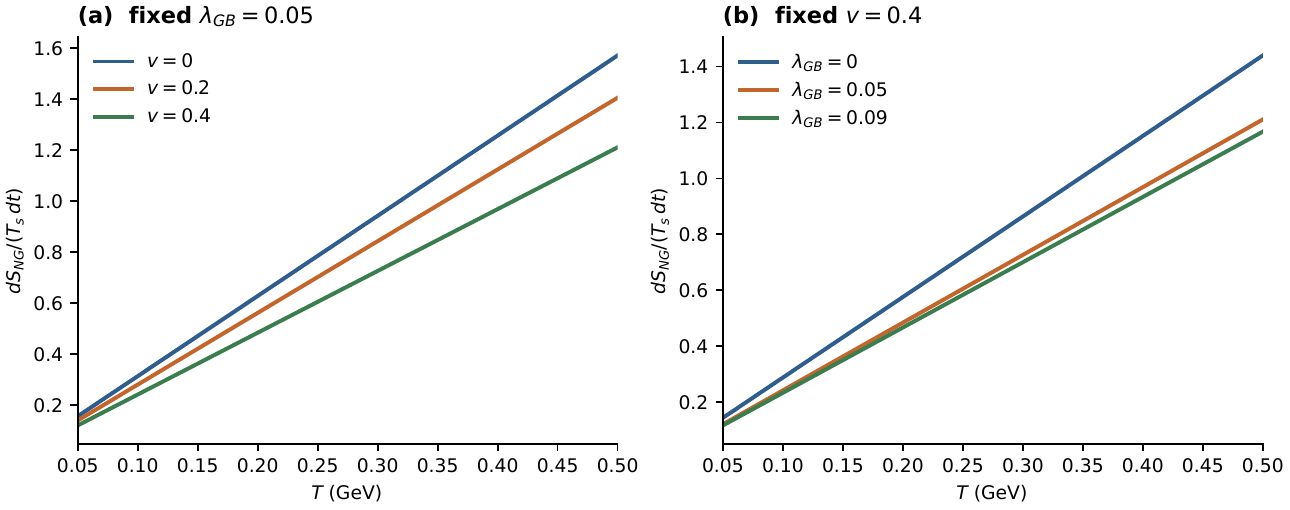}
    \caption{\label{fig3} \rev{Action growth versus temperature. Panel (a) shows different string velocities at $\lambda_{GB}=0.05$, while panel (b) shows different $\lambda_{GB}$ at $v=0.4$.}}
\end{figure}

\rev{Finally, we examine the temperature dependence of the NG action growth in Fig.~\ref{fig3}.}
\rev{The numerical curves are linear in temperature $T$. This follows from the relation $z_h=\sqrt{a}/(\pi T)$, after rescaling the radial coordinate by $z_h$, Eq.~(\ref{eqfn}) factorizes into an overall temperature scale times a dimensionless function of $v$ and $\lambda_{GB}$. The static result in Eq.~(\ref{eqstatic}) is the simplest example, giving $dS_{NG}/dt=T_sL^2\pi T$. For moving strings, the same linear temperature dependence remains, while the slope is reduced by increasing $|v|$ or by increasing $\lambda_{GB}$. Therefore, a higher temperature enhances the thermal state complexity growth, whereas motion and GB corrections reduce the proportionality coefficient.}

\section{Conclusion and discussion}\label{sec:04}
In this study, we investigate the holographic complexity growth in GB gravity under the CA conjecture, focusing on the influence of GB corrections characterized by the coupling parameter $\lambda_{GB}$. 
By introducing a probe string into the $(4+1)$-dimensional GB $AdS$ black brane background, we studied the time derivative of the NG action, which serves as a dual quantity to the complexity growth.
Our numerical analysis investigated the influence of $\lambda_{GB}$, string velocity, and temperature on the growth of holographic complexity. The main findings are summarized below.

The complexity growth exhibits a symmetric dependence on the string velocity, implying that it is sensitive to the magnitude of the velocity rather than its direction.
This symmetry arises from the isotropy of the system, which constrains that the NG action depends exclusively on the quadratic term in $v$.
The complexity growth reaches its maximum when the string is stationary and decreases monotonically with increasing string speed, indicating that the motion of the probe string suppresses complexity growth.
\rev{The stationary string growth rate is protected by the temperature relation $z_h=\sqrt{a}/(\pi T)$ and is exactly independent of $\lambda_{GB}$ at fixed temperature. For moving strings, $\lambda_{GB}$ suppresses the NG action growth. Thus the GB correction either leaves the static result unchanged or reduces the moving string contribution.}
\rev{The NG action growth is proportional to the temperature at fixed $v$ and $\lambda_{GB}$. Thus increasing the temperature strengthens the thermal contribution to complexification, while velocity and $\lambda_{GB}$ reduce the slope of this linear dependence.}

\rev{Several extensions would be useful for clarifying the physical meaning of the present result. First, it would be natural to study $R^4$ corrections, which represent higher-order stringy effects and may reveal whether the suppression found here is a generic property of finite coupling corrections. Second, comparing the NG action growth with other probe string observables, such as the drag force or jet quenching parameter, may help distinguish the complexity growth of a moving nonlocal probe from ordinary dissipative energy loss. Finally, it would be interesting to include charge, chemical potential, or flavor backreaction to test whether the linear temperature dependence and the GB suppression persist in more QCD like holographic backgrounds.}


\begin{thebibliography}{49}%
\makeatletter
\providecommand \@ifxundefined [1]{%
 \@ifx{#1\undefined}
}%
\providecommand \@ifnum [1]{%
 \ifnum #1\expandafter \@firstoftwo
 \else \expandafter \@secondoftwo
 \fi
}%
\providecommand \@ifx [1]{%
 \ifx #1\expandafter \@firstoftwo
 \else \expandafter \@secondoftwo
 \fi
}%
\providecommand \natexlab [1]{#1}%
\providecommand \enquote  [1]{``#1''}%
\providecommand \bibnamefont  [1]{#1}%
\providecommand \bibfnamefont [1]{#1}%
\providecommand \citenamefont [1]{#1}%
\providecommand \href@noop [0]{\@secondoftwo}%
\providecommand \href [0]{\begingroup \@sanitize@url \@href}%
\providecommand \@href[1]{\@@startlink{#1}\@@href}%
\providecommand \@@href[1]{\endgroup#1\@@endlink}%
\providecommand \@sanitize@url [0]{\catcode `\\12\catcode `\$12\catcode `\&12\catcode `\#12\catcode `\^12\catcode `\_12\catcode `\%12\relax}%
\providecommand \@@startlink[1]{}%
\providecommand \@@endlink[0]{}%
\providecommand \url  [0]{\begingroup\@sanitize@url \@url }%
\providecommand \@url [1]{\endgroup\@href {#1}{\urlprefix }}%
\providecommand \urlprefix  [0]{URL }%
\providecommand \Eprint [0]{\href }%
\providecommand \doibase [0]{http://dx.doi.org/}%
\providecommand \selectlanguage [0]{\@gobble}%
\providecommand \bibinfo  [0]{\@secondoftwo}%
\providecommand \bibfield  [0]{\@secondoftwo}%
\providecommand \translation [1]{[#1]}%
\providecommand \BibitemOpen [0]{}%
\providecommand \bibitemStop [0]{}%
\providecommand \bibitemNoStop [0]{.\EOS\space}%
\providecommand \EOS [0]{\spacefactor3000\relax}%
\providecommand \BibitemShut  [1]{\csname bibitem#1\endcsname}%
\let\auto@bib@innerbib\@empty
\bibitem [{\citenamefont {Watrous}(2008)}]{Watrous:2008any}%
  \BibitemOpen
  \bibfield  {author} {\bibinfo {author} {\bibfnamefont {J.}~\bibnamefont {Watrous}},\ }\href {\doibase 10.1007/978-0-387-30440-3{\_}428} {\  (\bibinfo {year} {2008}),\ 10.1007/978-0-387-30440-3{\_}428},\ \Eprint {http://arxiv.org/abs/0804.3401} {arXiv:0804.3401 [quant-ph]} \BibitemShut {NoStop}%
\bibitem [{\citenamefont {Osborne}(2012)}]{Osborne_2012}%
  \BibitemOpen
  \bibfield  {author} {\bibinfo {author} {\bibfnamefont {T.~J.}\ \bibnamefont {Osborne}},\ }\href {\doibase 10.1088/0034-4885/75/2/022001} {\bibfield  {journal} {\bibinfo  {journal} {Reports on Progress in Physics}\ }\textbf {\bibinfo {volume} {75}},\ \bibinfo {pages} {022001} (\bibinfo {year} {2012})}\BibitemShut {NoStop}%
\bibitem [{\citenamefont {Swingle}\ \emph {et~al.}(2016)\citenamefont {Swingle}, \citenamefont {Bentsen}, \citenamefont {Schleier-Smith},\ and\ \citenamefont {Hayden}}]{PhysRevA.94.040302}%
  \BibitemOpen
  \bibfield  {author} {\bibinfo {author} {\bibfnamefont {B.}~\bibnamefont {Swingle}}, \bibinfo {author} {\bibfnamefont {G.}~\bibnamefont {Bentsen}}, \bibinfo {author} {\bibfnamefont {M.}~\bibnamefont {Schleier-Smith}}, \ and\ \bibinfo {author} {\bibfnamefont {P.}~\bibnamefont {Hayden}},\ }\href {\doibase 10.1103/PhysRevA.94.040302} {\bibfield  {journal} {\bibinfo  {journal} {Phys. Rev. A}\ }\textbf {\bibinfo {volume} {94}},\ \bibinfo {pages} {040302} (\bibinfo {year} {2016})}\BibitemShut {NoStop}%
\bibitem [{\citenamefont {Banerjee}\ \emph {et~al.}(2018)\citenamefont {Banerjee}, \citenamefont {Erdmenger},\ and\ \citenamefont {Sarkar}}]{Banerjee:2017qti}%
  \BibitemOpen
  \bibfield  {author} {\bibinfo {author} {\bibfnamefont {S.}~\bibnamefont {Banerjee}}, \bibinfo {author} {\bibfnamefont {J.}~\bibnamefont {Erdmenger}}, \ and\ \bibinfo {author} {\bibfnamefont {D.}~\bibnamefont {Sarkar}},\ }\href {\doibase 10.1007/JHEP08(2018)001} {\bibfield  {journal} {\bibinfo  {journal} {JHEP}\ }\textbf {\bibinfo {volume} {08}},\ \bibinfo {pages} {001} (\bibinfo {year} {2018})},\ \Eprint {http://arxiv.org/abs/1701.02319} {arXiv:1701.02319 [hep-th]} \BibitemShut {NoStop}%
\bibitem [{\citenamefont {Takayanagi}(2012)}]{Takayanagi:2012kg}%
  \BibitemOpen
  \bibfield  {author} {\bibinfo {author} {\bibfnamefont {T.}~\bibnamefont {Takayanagi}},\ }\href {\doibase 10.1088/0264-9381/29/15/153001} {\bibfield  {journal} {\bibinfo  {journal} {Class. Quant. Grav.}\ }\textbf {\bibinfo {volume} {29}},\ \bibinfo {pages} {153001} (\bibinfo {year} {2012})},\ \Eprint {http://arxiv.org/abs/1204.2450} {arXiv:1204.2450 [gr-qc]} \BibitemShut {NoStop}%
\bibitem [{\citenamefont {Hubeny}\ \emph {et~al.}(2007)\citenamefont {Hubeny}, \citenamefont {Rangamani},\ and\ \citenamefont {Takayanagi}}]{Hubeny:2007xt}%
  \BibitemOpen
  \bibfield  {author} {\bibinfo {author} {\bibfnamefont {V.~E.}\ \bibnamefont {Hubeny}}, \bibinfo {author} {\bibfnamefont {M.}~\bibnamefont {Rangamani}}, \ and\ \bibinfo {author} {\bibfnamefont {T.}~\bibnamefont {Takayanagi}},\ }\href {\doibase 10.1088/1126-6708/2007/07/062} {\bibfield  {journal} {\bibinfo  {journal} {JHEP}\ }\textbf {\bibinfo {volume} {07}},\ \bibinfo {pages} {062} (\bibinfo {year} {2007})},\ \Eprint {http://arxiv.org/abs/0705.0016} {arXiv:0705.0016 [hep-th]} \BibitemShut {NoStop}%
\bibitem [{\citenamefont {Zhao}(2018)}]{Zhao:2017iul}%
  \BibitemOpen
  \bibfield  {author} {\bibinfo {author} {\bibfnamefont {Y.}~\bibnamefont {Zhao}},\ }\href {\doibase 10.1103/PhysRevD.98.086011} {\bibfield  {journal} {\bibinfo  {journal} {Phys. Rev. D}\ }\textbf {\bibinfo {volume} {98}},\ \bibinfo {pages} {086011} (\bibinfo {year} {2018})},\ \Eprint {http://arxiv.org/abs/1702.03957} {arXiv:1702.03957 [hep-th]} \BibitemShut {NoStop}%
\bibitem [{\citenamefont {Cazares}(2013)}]{Cazares:2012gxg}%
  \BibitemOpen
  \bibfield  {author} {\bibinfo {author} {\bibfnamefont {J.~A.}\ \bibnamefont {Cazares}},\ }\href {\doibase 10.1007/JHEP06(2013)001} {\bibfield  {journal} {\bibinfo  {journal} {JHEP}\ }\textbf {\bibinfo {volume} {06}},\ \bibinfo {pages} {001} (\bibinfo {year} {2013})},\ \Eprint {http://arxiv.org/abs/1211.6685} {arXiv:1211.6685 [hep-th]} \BibitemShut {NoStop}%
\bibitem [{\citenamefont {Hayden}\ and\ \citenamefont {Preskill}(2007)}]{Hayden:2007cs}%
  \BibitemOpen
  \bibfield  {author} {\bibinfo {author} {\bibfnamefont {P.}~\bibnamefont {Hayden}}\ and\ \bibinfo {author} {\bibfnamefont {J.}~\bibnamefont {Preskill}},\ }\href {\doibase 10.1088/1126-6708/2007/09/120} {\bibfield  {journal} {\bibinfo  {journal} {JHEP}\ }\textbf {\bibinfo {volume} {09}},\ \bibinfo {pages} {120} (\bibinfo {year} {2007})},\ \Eprint {http://arxiv.org/abs/0708.4025} {arXiv:0708.4025 [hep-th]} \BibitemShut {NoStop}%
\bibitem [{\citenamefont {Susskind}(2016)}]{Susskind:2015toa}%
  \BibitemOpen
  \bibfield  {author} {\bibinfo {author} {\bibfnamefont {L.}~\bibnamefont {Susskind}},\ }\href {\doibase 10.1002/prop.201500091} {\bibfield  {journal} {\bibinfo  {journal} {Fortsch. Phys.}\ }\textbf {\bibinfo {volume} {64}},\ \bibinfo {pages} {84} (\bibinfo {year} {2016})},\ \Eprint {http://arxiv.org/abs/1507.02287} {arXiv:1507.02287 [hep-th]} \BibitemShut {NoStop}%
\bibitem [{\citenamefont {Stanford}\ and\ \citenamefont {Susskind}(2014)}]{Stanford:2014jda}%
  \BibitemOpen
  \bibfield  {author} {\bibinfo {author} {\bibfnamefont {D.}~\bibnamefont {Stanford}}\ and\ \bibinfo {author} {\bibfnamefont {L.}~\bibnamefont {Susskind}},\ }\href {\doibase 10.1103/PhysRevD.90.126007} {\bibfield  {journal} {\bibinfo  {journal} {Phys. Rev. D}\ }\textbf {\bibinfo {volume} {90}},\ \bibinfo {pages} {126007} (\bibinfo {year} {2014})},\ \Eprint {http://arxiv.org/abs/1406.2678} {arXiv:1406.2678 [hep-th]} \BibitemShut {NoStop}%
\bibitem [{\citenamefont {Brown}\ \emph {et~al.}(2016{\natexlab{a}})\citenamefont {Brown}, \citenamefont {Roberts}, \citenamefont {Susskind}, \citenamefont {Swingle},\ and\ \citenamefont {Zhao}}]{Brown:2015bva}%
  \BibitemOpen
  \bibfield  {author} {\bibinfo {author} {\bibfnamefont {A.~R.}\ \bibnamefont {Brown}}, \bibinfo {author} {\bibfnamefont {D.~A.}\ \bibnamefont {Roberts}}, \bibinfo {author} {\bibfnamefont {L.}~\bibnamefont {Susskind}}, \bibinfo {author} {\bibfnamefont {B.}~\bibnamefont {Swingle}}, \ and\ \bibinfo {author} {\bibfnamefont {Y.}~\bibnamefont {Zhao}},\ }\href {\doibase 10.1103/PhysRevLett.116.191301} {\bibfield  {journal} {\bibinfo  {journal} {Phys. Rev. Lett.}\ }\textbf {\bibinfo {volume} {116}},\ \bibinfo {pages} {191301} (\bibinfo {year} {2016}{\natexlab{a}})},\ \Eprint {http://arxiv.org/abs/1509.07876} {arXiv:1509.07876 [hep-th]} \BibitemShut {NoStop}%
\bibitem [{\citenamefont {Brown}\ \emph {et~al.}(2016{\natexlab{b}})\citenamefont {Brown}, \citenamefont {Roberts}, \citenamefont {Susskind}, \citenamefont {Swingle},\ and\ \citenamefont {Zhao}}]{Brown:2015lvg}%
  \BibitemOpen
  \bibfield  {author} {\bibinfo {author} {\bibfnamefont {A.~R.}\ \bibnamefont {Brown}}, \bibinfo {author} {\bibfnamefont {D.~A.}\ \bibnamefont {Roberts}}, \bibinfo {author} {\bibfnamefont {L.}~\bibnamefont {Susskind}}, \bibinfo {author} {\bibfnamefont {B.}~\bibnamefont {Swingle}}, \ and\ \bibinfo {author} {\bibfnamefont {Y.}~\bibnamefont {Zhao}},\ }\href {\doibase 10.1103/PhysRevD.93.086006} {\bibfield  {journal} {\bibinfo  {journal} {Phys. Rev. D}\ }\textbf {\bibinfo {volume} {93}},\ \bibinfo {pages} {086006} (\bibinfo {year} {2016}{\natexlab{b}})},\ \Eprint {http://arxiv.org/abs/1512.04993} {arXiv:1512.04993 [hep-th]} \BibitemShut {NoStop}%
\bibitem [{\citenamefont {Cai}\ \emph {et~al.}(2016)\citenamefont {Cai}, \citenamefont {Ruan}, \citenamefont {Wang}, \citenamefont {Yang},\ and\ \citenamefont {Peng}}]{Cai:2016xho}%
  \BibitemOpen
  \bibfield  {author} {\bibinfo {author} {\bibfnamefont {R.-G.}\ \bibnamefont {Cai}}, \bibinfo {author} {\bibfnamefont {S.-M.}\ \bibnamefont {Ruan}}, \bibinfo {author} {\bibfnamefont {S.-J.}\ \bibnamefont {Wang}}, \bibinfo {author} {\bibfnamefont {R.-Q.}\ \bibnamefont {Yang}}, \ and\ \bibinfo {author} {\bibfnamefont {R.-H.}\ \bibnamefont {Peng}},\ }\href {\doibase 10.1007/JHEP09(2016)161} {\bibfield  {journal} {\bibinfo  {journal} {JHEP}\ }\textbf {\bibinfo {volume} {09}},\ \bibinfo {pages} {161} (\bibinfo {year} {2016})},\ \Eprint {http://arxiv.org/abs/1606.08307} {arXiv:1606.08307 [gr-qc]} \BibitemShut {NoStop}%
\bibitem [{\citenamefont {Carmi}\ \emph {et~al.}(2017)\citenamefont {Carmi}, \citenamefont {Chapman}, \citenamefont {Marrochio}, \citenamefont {Myers},\ and\ \citenamefont {Sugishita}}]{Carmi:2017jqz}%
  \BibitemOpen
  \bibfield  {author} {\bibinfo {author} {\bibfnamefont {D.}~\bibnamefont {Carmi}}, \bibinfo {author} {\bibfnamefont {S.}~\bibnamefont {Chapman}}, \bibinfo {author} {\bibfnamefont {H.}~\bibnamefont {Marrochio}}, \bibinfo {author} {\bibfnamefont {R.~C.}\ \bibnamefont {Myers}}, \ and\ \bibinfo {author} {\bibfnamefont {S.}~\bibnamefont {Sugishita}},\ }\href {\doibase 10.1007/JHEP11(2017)188} {\bibfield  {journal} {\bibinfo  {journal} {JHEP}\ }\textbf {\bibinfo {volume} {11}},\ \bibinfo {pages} {188} (\bibinfo {year} {2017})},\ \Eprint {http://arxiv.org/abs/1709.10184} {arXiv:1709.10184 [hep-th]} \BibitemShut {NoStop}%
\bibitem [{\citenamefont {Couch}\ \emph {et~al.}(2017)\citenamefont {Couch}, \citenamefont {Fischler},\ and\ \citenamefont {Nguyen}}]{Couch:2016exn}%
  \BibitemOpen
  \bibfield  {author} {\bibinfo {author} {\bibfnamefont {J.}~\bibnamefont {Couch}}, \bibinfo {author} {\bibfnamefont {W.}~\bibnamefont {Fischler}}, \ and\ \bibinfo {author} {\bibfnamefont {P.~H.}\ \bibnamefont {Nguyen}},\ }\href {\doibase 10.1007/JHEP03(2017)119} {\bibfield  {journal} {\bibinfo  {journal} {JHEP}\ }\textbf {\bibinfo {volume} {03}},\ \bibinfo {pages} {119} (\bibinfo {year} {2017})},\ \Eprint {http://arxiv.org/abs/1610.02038} {arXiv:1610.02038 [hep-th]} \BibitemShut {NoStop}%
\bibitem [{\citenamefont {Belin}\ \emph {et~al.}(2022)\citenamefont {Belin}, \citenamefont {Myers}, \citenamefont {Ruan}, \citenamefont {S{\'a}rosi},\ and\ \citenamefont {Speranza}}]{Belin:2021bga}%
  \BibitemOpen
  \bibfield  {author} {\bibinfo {author} {\bibfnamefont {A.}~\bibnamefont {Belin}}, \bibinfo {author} {\bibfnamefont {R.~C.}\ \bibnamefont {Myers}}, \bibinfo {author} {\bibfnamefont {S.-M.}\ \bibnamefont {Ruan}}, \bibinfo {author} {\bibfnamefont {G.}~\bibnamefont {S{\'a}rosi}}, \ and\ \bibinfo {author} {\bibfnamefont {A.~J.}\ \bibnamefont {Speranza}},\ }\href {\doibase 10.1103/PhysRevLett.128.081602} {\bibfield  {journal} {\bibinfo  {journal} {Phys. Rev. Lett.}\ }\textbf {\bibinfo {volume} {128}},\ \bibinfo {pages} {081602} (\bibinfo {year} {2022})},\ \Eprint {http://arxiv.org/abs/2111.02429} {arXiv:2111.02429 [hep-th]} \BibitemShut {NoStop}%
\bibitem [{\citenamefont {Belin}\ \emph {et~al.}(2023)\citenamefont {Belin}, \citenamefont {Myers}, \citenamefont {Ruan}, \citenamefont {S{\'a}rosi},\ and\ \citenamefont {Speranza}}]{Belin:2022xmt}%
  \BibitemOpen
  \bibfield  {author} {\bibinfo {author} {\bibfnamefont {A.}~\bibnamefont {Belin}}, \bibinfo {author} {\bibfnamefont {R.~C.}\ \bibnamefont {Myers}}, \bibinfo {author} {\bibfnamefont {S.-M.}\ \bibnamefont {Ruan}}, \bibinfo {author} {\bibfnamefont {G.}~\bibnamefont {S{\'a}rosi}}, \ and\ \bibinfo {author} {\bibfnamefont {A.~J.}\ \bibnamefont {Speranza}},\ }\href {\doibase 10.1007/JHEP01(2023)154} {\bibfield  {journal} {\bibinfo  {journal} {JHEP}\ }\textbf {\bibinfo {volume} {01}},\ \bibinfo {pages} {154} (\bibinfo {year} {2023})},\ \Eprint {http://arxiv.org/abs/2210.09647} {arXiv:2210.09647 [hep-th]} \BibitemShut {NoStop}%
\bibitem [{\citenamefont {J{\o}rstad}\ \emph {et~al.}(2023)\citenamefont {J{\o}rstad}, \citenamefont {Myers},\ and\ \citenamefont {Ruan}}]{Jorstad:2023kmq}%
  \BibitemOpen
  \bibfield  {author} {\bibinfo {author} {\bibfnamefont {E.}~\bibnamefont {J{\o}rstad}}, \bibinfo {author} {\bibfnamefont {R.~C.}\ \bibnamefont {Myers}}, \ and\ \bibinfo {author} {\bibfnamefont {S.-M.}\ \bibnamefont {Ruan}},\ }\href {\doibase 10.1007/JHEP07(2023)223} {\bibfield  {journal} {\bibinfo  {journal} {JHEP}\ }\textbf {\bibinfo {volume} {07}},\ \bibinfo {pages} {223} (\bibinfo {year} {2023})},\ \Eprint {http://arxiv.org/abs/2304.05453} {arXiv:2304.05453 [hep-th]} \BibitemShut {NoStop}%
\bibitem [{\citenamefont {Myers}\ and\ \citenamefont {Ruan}(2024)}]{Myers:2024vve}%
  \BibitemOpen
  \bibfield  {author} {\bibinfo {author} {\bibfnamefont {R.~C.}\ \bibnamefont {Myers}}\ and\ \bibinfo {author} {\bibfnamefont {S.-M.}\ \bibnamefont {Ruan}}\ }(\bibinfo {year} {2024})\ \Eprint {http://arxiv.org/abs/2403.17475} {arXiv:2403.17475 [hep-th]} \BibitemShut {NoStop}%
\bibitem [{\citenamefont {Moosa}(2018)}]{Moosa:2017yvt}%
  \BibitemOpen
  \bibfield  {author} {\bibinfo {author} {\bibfnamefont {M.}~\bibnamefont {Moosa}},\ }\href {\doibase 10.1007/JHEP03(2018)031} {\bibfield  {journal} {\bibinfo  {journal} {JHEP}\ }\textbf {\bibinfo {volume} {03}},\ \bibinfo {pages} {031} (\bibinfo {year} {2018})},\ \Eprint {http://arxiv.org/abs/1711.02668} {arXiv:1711.02668 [hep-th]} \BibitemShut {NoStop}%
\bibitem [{\citenamefont {Abad}\ \emph {et~al.}(2018)\citenamefont {Abad}, \citenamefont {Kulaxizi},\ and\ \citenamefont {Parnachev}}]{Abad:2017cgl}%
  \BibitemOpen
  \bibfield  {author} {\bibinfo {author} {\bibfnamefont {F.~J.~G.}\ \bibnamefont {Abad}}, \bibinfo {author} {\bibfnamefont {M.}~\bibnamefont {Kulaxizi}}, \ and\ \bibinfo {author} {\bibfnamefont {A.}~\bibnamefont {Parnachev}},\ }\href {\doibase 10.1007/JHEP01(2018)127} {\bibfield  {journal} {\bibinfo  {journal} {JHEP}\ }\textbf {\bibinfo {volume} {01}},\ \bibinfo {pages} {127} (\bibinfo {year} {2018})},\ \Eprint {http://arxiv.org/abs/1705.08424} {arXiv:1705.08424 [hep-th]} \BibitemShut {NoStop}%
\bibitem [{\citenamefont {Fu}\ \emph {et~al.}(2018)\citenamefont {Fu}, \citenamefont {Maloney}, \citenamefont {Marolf}, \citenamefont {Maxfield},\ and\ \citenamefont {Wang}}]{Fu:2018kcp}%
  \BibitemOpen
  \bibfield  {author} {\bibinfo {author} {\bibfnamefont {Z.}~\bibnamefont {Fu}}, \bibinfo {author} {\bibfnamefont {A.}~\bibnamefont {Maloney}}, \bibinfo {author} {\bibfnamefont {D.}~\bibnamefont {Marolf}}, \bibinfo {author} {\bibfnamefont {H.}~\bibnamefont {Maxfield}}, \ and\ \bibinfo {author} {\bibfnamefont {Z.}~\bibnamefont {Wang}},\ }\href {\doibase 10.1007/JHEP02(2018)072} {\bibfield  {journal} {\bibinfo  {journal} {JHEP}\ }\textbf {\bibinfo {volume} {02}},\ \bibinfo {pages} {072} (\bibinfo {year} {2018})},\ \Eprint {http://arxiv.org/abs/1801.01137} {arXiv:1801.01137 [hep-th]} \BibitemShut {NoStop}%
\bibitem [{\citenamefont {Nagasaki}(2017)}]{Nagasaki:2017kqe}%
  \BibitemOpen
  \bibfield  {author} {\bibinfo {author} {\bibfnamefont {K.}~\bibnamefont {Nagasaki}},\ }\href {\doibase 10.1103/PhysRevD.96.126018} {\bibfield  {journal} {\bibinfo  {journal} {Phys. Rev. D}\ }\textbf {\bibinfo {volume} {96}},\ \bibinfo {pages} {126018} (\bibinfo {year} {2017})},\ \Eprint {http://arxiv.org/abs/1707.08376} {arXiv:1707.08376 [hep-th]} \BibitemShut {NoStop}%
\bibitem [{\citenamefont {Nagasaki}(2018)}]{Nagasaki:2018csh}%
  \BibitemOpen
  \bibfield  {author} {\bibinfo {author} {\bibfnamefont {K.}~\bibnamefont {Nagasaki}},\ }\href {\doibase 10.1103/PhysRevD.98.126014} {\bibfield  {journal} {\bibinfo  {journal} {Phys. Rev. D}\ }\textbf {\bibinfo {volume} {98}},\ \bibinfo {pages} {126014} (\bibinfo {year} {2018})},\ \Eprint {http://arxiv.org/abs/1807.01088} {arXiv:1807.01088 [hep-th]} \BibitemShut {NoStop}%
\bibitem [{\citenamefont {An}\ \emph {et~al.}(2018)\citenamefont {An}, \citenamefont {Cai},\ and\ \citenamefont {Peng}}]{An:2018dbz}%
  \BibitemOpen
  \bibfield  {author} {\bibinfo {author} {\bibfnamefont {Y.-S.}\ \bibnamefont {An}}, \bibinfo {author} {\bibfnamefont {R.-G.}\ \bibnamefont {Cai}}, \ and\ \bibinfo {author} {\bibfnamefont {Y.}\ \bibnamefont {Peng}},\ }\href {\doibase 10.1103/PhysRevD.98.106013} {\bibfield  {journal} {\bibinfo  {journal} {Phys. Rev. D}\ }\textbf {\bibinfo {volume} {98}},\ \bibinfo {pages} {106013} (\bibinfo {year} {2018})},\ \Eprint {http://arxiv.org/abs/1805.07775} {arXiv:1805.07775 [hep-th]} \BibitemShut {NoStop}%
\bibitem [{\citenamefont {Nagasaki}(2020)}]{Nagasaki:2019icm}%
  \BibitemOpen
  \bibfield  {author} {\bibinfo {author} {\bibfnamefont {K.}~\bibnamefont {Nagasaki}},\ }\href {\doibase 10.1142/S0217751X20501523} {\bibfield  {journal} {\bibinfo  {journal} {Int. J. Mod. Phys. A}\ }\textbf {\bibinfo {volume} {35}},\ \bibinfo {pages} {2050152} (\bibinfo {year} {2020})},\ \Eprint {http://arxiv.org/abs/1912.03567} {arXiv:1912.03567 [hep-th]} \BibitemShut {NoStop}%
\bibitem [{\citenamefont {Nagasaki}(2023)}]{Nagasaki:2022lll}%
  \BibitemOpen
  \bibfield  {author} {\bibinfo {author} {\bibfnamefont {K.}~\bibnamefont {Nagasaki}},\ }\href {\doibase 10.1142/S0217751X23500276} {\bibfield  {journal} {\bibinfo  {journal} {Int. J. Mod. Phys. A}\ }\textbf {\bibinfo {volume} {38}},\ \bibinfo {pages} {2350027} (\bibinfo {year} {2023})},\ \Eprint {http://arxiv.org/abs/2205.00196} {arXiv:2205.00196 [hep-th]} \BibitemShut {NoStop}%
\bibitem [{\citenamefont {Zhou}\ \emph {et~al.}(2021)\citenamefont {Zhou}, \citenamefont {Kuang},\ and\ \citenamefont {Wu}}]{Zhou:2021vsm}%
  \BibitemOpen
  \bibfield  {author} {\bibinfo {author} {\bibfnamefont {Y.-T.}\ \bibnamefont {Zhou}}, \bibinfo {author} {\bibfnamefont {X.-M.}\ \bibnamefont {Kuang}}, \ and\ \bibinfo {author} {\bibfnamefont {J.-P.}\ \bibnamefont {Wu}},\ }\href {\doibase 10.1140/epjc/s10052-021-09563-1} {\bibfield  {journal} {\bibinfo  {journal} {Eur. Phys. J. C}\ }\textbf {\bibinfo {volume} {81}},\ \bibinfo {pages} {768} (\bibinfo {year} {2021})},\ \Eprint {http://arxiv.org/abs/2104.12998} {arXiv:2104.12998 [hep-th]} \BibitemShut {NoStop}%
\bibitem [{\citenamefont {Santos}(2020)}]{Santos:2020xox}%
  \BibitemOpen
  \bibfield  {author} {\bibinfo {author} {\bibfnamefont {F.~F.}\ \bibnamefont {Santos}},\ }\href {\doibase 10.1140/epjp/s13360-020-00805-x} {\bibfield  {journal} {\bibinfo  {journal} {Eur. Phys. J. Plus}\ }\textbf {\bibinfo {volume} {135}},\ \bibinfo {pages} {810} (\bibinfo {year} {2020})},\ \Eprint {http://arxiv.org/abs/2005.10983} {arXiv:2005.10983 [hep-th]} \BibitemShut {NoStop}%
\bibitem [{\citenamefont {Bravo-Gaete}\ and\ \citenamefont {Santos}(2022)}]{Bravo-Gaete:2020lzs}%
  \BibitemOpen
  \bibfield  {author} {\bibinfo {author} {\bibfnamefont {M.}~\bibnamefont {Bravo-Gaete}}\ and\ \bibinfo {author} {\bibfnamefont {F.~F.}\ \bibnamefont {Santos}},\ }\href {\doibase 10.1140/epjc/s10052-022-10064-y} {\bibfield  {journal} {\bibinfo  {journal} {Eur. Phys. J. C}\ }\textbf {\bibinfo {volume} {82}},\ \bibinfo {pages} {101} (\bibinfo {year} {2022})},\ \Eprint {http://arxiv.org/abs/2010.10942} {arXiv:2010.10942 [hep-th]} \BibitemShut {NoStop}%
\bibitem [{\citenamefont {Chang}\ and\ \citenamefont {Hou}(2024)}]{Chang:2024muq}%
  \BibitemOpen
  \bibfield  {author} {\bibinfo {author} {\bibfnamefont {W.-B.}\ \bibnamefont {Chang}}\ and\ \bibinfo {author} {\bibfnamefont {D.-f.}\ \bibnamefont {Hou}},\ }\href {\doibase 10.1088/1674-1137/ad1b3e} {\bibfield  {journal} {\bibinfo  {journal} {Chin. Phys. C}\ }\textbf {\bibinfo {volume} {48}},\ \bibinfo {pages} {034106} (\bibinfo {year} {2024})}\BibitemShut {NoStop}%
\bibitem [{\citenamefont {Chang}\ \emph {et~al.}(2025)\citenamefont {Chang}, \citenamefont {Chen},\ and\ \citenamefont {Hou}}]{Chang:2025flavor}%
  \BibitemOpen
  \bibfield  {author} {\bibinfo {author} {\bibfnamefont {W.-B.}\ \bibnamefont {Chang}}, \bibinfo {author} {\bibfnamefont {X.}~\bibnamefont {Chen}}, \ and\ \bibinfo {author} {\bibfnamefont {D.}~\bibnamefont {Hou}},\ }\bibfield  {title} {\enquote {\bibinfo {title} {Complexity Growth in Flavor-Dependent Systems}},\ }\bibinfo {year} {2025},\ \Eprint {http://arxiv.org/abs/2511.22799} {arXiv:2511.22799 [hep-ph]} \BibitemShut {NoStop}%
\bibitem [{\citenamefont {Guo}\ and\ \citenamefont {Zhang}(2025)}]{Guo:2025bgx}%
  \BibitemOpen
  \bibfield  {author} {\bibinfo {author} {\bibfnamefont {L.}~\bibnamefont {Guo}}\ and\ \bibinfo {author} {\bibfnamefont {Z.-q.}\ \bibnamefont {Zhang}},\ }\href {\doibase 10.1088/1674-1137/ada125} {\bibfield  {journal} {\bibinfo  {journal} {Chin. Phys. C}\ }\textbf {\bibinfo {volume} {49}},\ \bibinfo {pages} {035104} (\bibinfo {year} {2025})}\BibitemShut {NoStop}%
\bibitem [{\citenamefont {Guo}\ \emph {et~al.}(2017)\citenamefont {Guo}, \citenamefont {Wei}, \citenamefont {Li},\ and\ \citenamefont {Liu}}]{Guo:2017rul}%
  \BibitemOpen
  \bibfield  {author} {\bibinfo {author} {\bibfnamefont {W.-D.}\ \bibnamefont {Guo}}, \bibinfo {author} {\bibfnamefont {S.-W.}\ \bibnamefont {Wei}}, \bibinfo {author} {\bibfnamefont {Y.-Y.}\ \bibnamefont {Li}}, \ and\ \bibinfo {author} {\bibfnamefont {Y.-X.}\ \bibnamefont {Liu}},\ }\href {\doibase 10.1140/epjc/s10052-017-5466-5} {\bibfield  {journal} {\bibinfo  {journal} {Eur. Phys. J. C}\ }\textbf {\bibinfo {volume} {77}},\ \bibinfo {pages} {904} (\bibinfo {year} {2017})},\ \Eprint {http://arxiv.org/abs/1703.10468} {arXiv:1703.10468 [gr-qc]} \BibitemShut {NoStop}%
\bibitem [{\citenamefont {Aguilar-Gutierrez}\ \emph {et~al.}(2024{\natexlab{a}})\citenamefont {Aguilar-Gutierrez}, \citenamefont {Heller},\ and\ \citenamefont {Van~der Schueren}}]{Aguilar-Gutierrez:2023zqm}%
  \BibitemOpen
  \bibfield  {author} {\bibinfo {author} {\bibfnamefont {S.~E.}\ \bibnamefont {Aguilar-Gutierrez}}, \bibinfo {author} {\bibfnamefont {M.~P.}\ \bibnamefont {Heller}}, \ and\ \bibinfo {author} {\bibfnamefont {S.}~\bibnamefont {Van~der Schueren}},\ }\href {\doibase 10.1103/PhysRevD.110.066009} {\bibfield  {journal} {\bibinfo  {journal} {Phys. Rev. D}\ }\textbf {\bibinfo {volume} {110}},\ \bibinfo {pages} {066009} (\bibinfo {year} {2024}{\natexlab{a}})},\ \Eprint {http://arxiv.org/abs/2305.11280} {arXiv:2305.11280 [hep-th]} \BibitemShut {NoStop}%
\bibitem [{\citenamefont {Pedraza}\ \emph {et~al.}(2022)\citenamefont {Pedraza}, \citenamefont {Russo}, \citenamefont {Svesko},\ and\ \citenamefont {Weller-Davies}}]{Pedraza:2021fgp}%
  \BibitemOpen
  \bibfield  {author} {\bibinfo {author} {\bibfnamefont {J.~F.}\ \bibnamefont {Pedraza}}, \bibinfo {author} {\bibfnamefont {A.}~\bibnamefont {Russo}}, \bibinfo {author} {\bibfnamefont {A.}~\bibnamefont {Svesko}}, \ and\ \bibinfo {author} {\bibfnamefont {Z.}~\bibnamefont {Weller-Davies}},\ }\href {\doibase 10.1007/JHEP02(2022)093} {\bibfield  {journal} {\bibinfo  {journal} {JHEP}\ }\textbf {\bibinfo {volume} {02}},\ \bibinfo {pages} {093} (\bibinfo {year} {2022})},\ \Eprint {http://arxiv.org/abs/2106.12585} {arXiv:2106.12585 [hep-th]} \BibitemShut {NoStop}%
\bibitem [{\citenamefont {Chen}\ \emph {et~al.}(2024)\citenamefont {Chen}, \citenamefont {Liu},\ and\ \citenamefont {Yu}}]{Chen:2023tpi}%
  \BibitemOpen
  \bibfield  {author} {\bibinfo {author} {\bibfnamefont {B.}~\bibnamefont {Chen}}, \bibinfo {author} {\bibfnamefont {Y.}~\bibnamefont {Liu}}, \ and\ \bibinfo {author} {\bibfnamefont {B.}~\bibnamefont {Yu}},\ }\href {\doibase 10.1007/JHEP01(2024)055} {\bibfield  {journal} {\bibinfo  {journal} {JHEP}\ }\textbf {\bibinfo {volume} {01}},\ \bibinfo {pages} {055} (\bibinfo {year} {2024})},\ \Eprint {http://arxiv.org/abs/2307.15968} {arXiv:2307.15968 [hep-th]} \BibitemShut {NoStop}%
\bibitem [{\citenamefont {Aguilar-Gutierrez}\ \emph {et~al.}(2024{\natexlab{b}})\citenamefont {Aguilar-Gutierrez}, \citenamefont {Craps}, \citenamefont {Hernandez}, \citenamefont {Khramtsov}, \citenamefont {Knysh},\ and\ \citenamefont {Shukla}}]{Aguilar-Gutierrez:2023ccv}%
  \BibitemOpen
  \bibfield  {author} {\bibinfo {author} {\bibfnamefont {S.~E.}\ \bibnamefont {Aguilar-Gutierrez}}, \bibinfo {author} {\bibfnamefont {B.}~\bibnamefont {Craps}}, \bibinfo {author} {\bibfnamefont {J.}~\bibnamefont {Hernandez}}, \bibinfo {author} {\bibfnamefont {M.}~\bibnamefont {Khramtsov}}, \bibinfo {author} {\bibfnamefont {M.}~\bibnamefont {Knysh}}, \ and\ \bibinfo {author} {\bibfnamefont {A.}~\bibnamefont {Shukla}},\ }\href {\doibase 10.1007/JHEP03(2024)173} {\bibfield  {journal} {\bibinfo  {journal} {JHEP}\ }\textbf {\bibinfo {volume} {03}},\ \bibinfo {pages} {173} (\bibinfo {year} {2024}{\natexlab{b}})},\ \Eprint {http://arxiv.org/abs/2312.12349} {arXiv:2312.12349 [hep-th]} \BibitemShut {NoStop}%
\bibitem [{\citenamefont {C{\'a}ceres}\ \emph {et~al.}(2024)\citenamefont {C{\'a}ceres}, \citenamefont {Murcia}, \citenamefont {Patra},\ and\ \citenamefont {Pedraza}}]{Caceres:2024edr}%
  \BibitemOpen
  \bibfield  {author} {\bibinfo {author} {\bibfnamefont {E.}~\bibnamefont {C{\'a}ceres}}, \bibinfo {author} {\bibfnamefont {{\'A}.~J.}\ \bibnamefont {Murcia}}, \bibinfo {author} {\bibfnamefont {A.~K.}\ \bibnamefont {Patra}}, \ and\ \bibinfo {author} {\bibfnamefont {J.~F.}\ \bibnamefont {Pedraza}},\ }\href {\doibase 10.1007/JHEP12(2024)077} {\bibfield  {journal} {\bibinfo  {journal} {JHEP}\ }\textbf {\bibinfo {volume} {12}},\ \bibinfo {pages} {077} (\bibinfo {year} {2024})},\ \Eprint {http://arxiv.org/abs/2408.14535} {arXiv:2408.14535 [hep-th]} \BibitemShut {NoStop}%
\bibitem [{\citenamefont {Jiang}\ and\ \citenamefont {Liu}(2025)}]{Jiang:2025qai}%
  \BibitemOpen
  \bibfield  {author} {\bibinfo {author} {\bibfnamefont {H.-Y.}\ \bibnamefont {Jiang}}\ and\ \bibinfo {author} {\bibfnamefont {Y.-X.}\ \bibnamefont {Liu}},\ }\href@noop {} {\  (\bibinfo {year} {2025})},\ \Eprint {http://arxiv.org/abs/2506.10398} {arXiv:2506.10398 [hep-th]} \BibitemShut {NoStop}%
\bibitem [{\citenamefont {Douglas}\ and\ \citenamefont {Kachru}(2007)}]{Douglas:2006es}%
  \BibitemOpen
  \bibfield  {author} {\bibinfo {author} {\bibfnamefont {M.~R.}\ \bibnamefont {Douglas}}\ and\ \bibinfo {author} {\bibfnamefont {S.}~\bibnamefont {Kachru}},\ }\href {\doibase 10.1103/RevModPhys.79.733} {\bibfield  {journal} {\bibinfo  {journal} {Rev. Mod. Phys.}\ }\textbf {\bibinfo {volume} {79}},\ \bibinfo {pages} {733} (\bibinfo {year} {2007})},\ \Eprint {http://arxiv.org/abs/hep-th/0610102} {arXiv:hep-th/0610102} \BibitemShut {NoStop}%
\bibitem [{\citenamefont {Buchel}\ \emph {et~al.}(2009)\citenamefont {Buchel}, \citenamefont {Myers},\ and\ \citenamefont {Sinha}}]{Buchel:2008vz}%
  \BibitemOpen
  \bibfield  {author} {\bibinfo {author} {\bibfnamefont {A.}~\bibnamefont {Buchel}}, \bibinfo {author} {\bibfnamefont {R.~C.}\ \bibnamefont {Myers}}, \ and\ \bibinfo {author} {\bibfnamefont {A.}~\bibnamefont {Sinha}},\ }\href {\doibase 10.1088/1126-6708/2009/03/084} {\bibfield  {journal} {\bibinfo  {journal} {JHEP}\ }\textbf {\bibinfo {volume} {03}},\ \bibinfo {pages} {084} (\bibinfo {year} {2009})},\ \Eprint {http://arxiv.org/abs/0812.2521} {arXiv:0812.2521 [hep-th]} \BibitemShut {NoStop}%
\bibitem [{\citenamefont {Aharony}\ \emph {et~al.}(1999)\citenamefont {Aharony}, \citenamefont {Pawelczyk}, \citenamefont {Theisen},\ and\ \citenamefont {Yankielowicz}}]{Aharony:1999rz}%
  \BibitemOpen
  \bibfield  {author} {\bibinfo {author} {\bibfnamefont {O.}~\bibnamefont {Aharony}}, \bibinfo {author} {\bibfnamefont {J.}~\bibnamefont {Pawelczyk}}, \bibinfo {author} {\bibfnamefont {S.}~\bibnamefont {Theisen}}, \ and\ \bibinfo {author} {\bibfnamefont {S.}~\bibnamefont {Yankielowicz}},\ }\href {\doibase 10.1103/PhysRevD.60.066001} {\bibfield  {journal} {\bibinfo  {journal} {Phys. Rev. D}\ }\textbf {\bibinfo {volume} {60}},\ \bibinfo {pages} {066001} (\bibinfo {year} {1999})},\ \Eprint {http://arxiv.org/abs/hep-th/9901134} {arXiv:hep-th/9901134} \BibitemShut {NoStop}%
\bibitem [{\citenamefont {Aharony}\ and\ \citenamefont {Tachikawa}(2008)}]{Aharony:2007dj}%
  \BibitemOpen
  \bibfield  {author} {\bibinfo {author} {\bibfnamefont {O.}~\bibnamefont {Aharony}}\ and\ \bibinfo {author} {\bibfnamefont {Y.}~\bibnamefont {Tachikawa}},\ }\href {\doibase 10.1088/1126-6708/2008/01/037} {\bibfield  {journal} {\bibinfo  {journal} {JHEP}\ }\textbf {\bibinfo {volume} {01}},\ \bibinfo {pages} {037} (\bibinfo {year} {2008})},\ \Eprint {http://arxiv.org/abs/0711.4532} {arXiv:0711.4532 [hep-th]} \BibitemShut {NoStop}%
\bibitem [{\citenamefont {Brigante}\ \emph {et~al.}(2008{\natexlab{a}})\citenamefont {Brigante}, \citenamefont {Liu}, \citenamefont {Myers}, \citenamefont {Shenker},\ and\ \citenamefont {Yaida}}]{Brigante:2007nu}%
  \BibitemOpen
  \bibfield  {author} {\bibinfo {author} {\bibfnamefont {M.}~\bibnamefont {Brigante}}, \bibinfo {author} {\bibfnamefont {H.}~\bibnamefont {Liu}}, \bibinfo {author} {\bibfnamefont {R.~C.}\ \bibnamefont {Myers}}, \bibinfo {author} {\bibfnamefont {S.}~\bibnamefont {Shenker}}, \ and\ \bibinfo {author} {\bibfnamefont {S.}~\bibnamefont {Yaida}},\ }\href {\doibase 10.1103/PhysRevD.77.126006} {\bibfield  {journal} {\bibinfo  {journal} {Phys. Rev. D}\ }\textbf {\bibinfo {volume} {77}},\ \bibinfo {pages} {126006} (\bibinfo {year} {2008}{\natexlab{a}})},\ \Eprint {http://arxiv.org/abs/0712.0805} {arXiv:0712.0805 [hep-th]} \BibitemShut {NoStop}%
\bibitem [{\citenamefont {Brigante}\ \emph {et~al.}(2008{\natexlab{b}})\citenamefont {Brigante}, \citenamefont {Liu}, \citenamefont {Myers}, \citenamefont {Shenker},\ and\ \citenamefont {Yaida}}]{Brigante:2008gz}%
  \BibitemOpen
  \bibfield  {author} {\bibinfo {author} {\bibfnamefont {M.}~\bibnamefont {Brigante}}, \bibinfo {author} {\bibfnamefont {H.}~\bibnamefont {Liu}}, \bibinfo {author} {\bibfnamefont {R.~C.}\ \bibnamefont {Myers}}, \bibinfo {author} {\bibfnamefont {S.}~\bibnamefont {Shenker}}, \ and\ \bibinfo {author} {\bibfnamefont {S.}~\bibnamefont {Yaida}},\ }\href {\doibase 10.1103/PhysRevLett.100.191601} {\bibfield  {journal} {\bibinfo  {journal} {Phys. Rev. Lett.}\ }\textbf {\bibinfo {volume} {100}},\ \bibinfo {pages} {191601} (\bibinfo {year} {2008}{\natexlab{b}})},\ \Eprint {http://arxiv.org/abs/0802.3318} {arXiv:0802.3318 [hep-th]} \BibitemShut {NoStop}%
\bibitem [{\citenamefont {Zwiebach}(1985)}]{Zwiebach:1985uq}%
  \BibitemOpen
  \bibfield  {author} {\bibinfo {author} {\bibfnamefont {B.}~\bibnamefont {Zwiebach}},\ }\href {\doibase 10.1016/0370-2693(85)91616-8} {\bibfield  {journal} {\bibinfo  {journal} {Phys. Lett. B}\ }\textbf {\bibinfo {volume} {156}},\ \bibinfo {pages} {315} (\bibinfo {year} {1985})}\BibitemShut {NoStop}%
\bibitem [{\citenamefont {Cai}(2002)}]{Cai:2001dz}%
  \BibitemOpen
  \bibfield  {author} {\bibinfo {author} {\bibfnamefont {R.-G.}\ \bibnamefont {Cai}},\ }\href {\doibase 10.1103/PhysRevD.65.084014} {\bibfield  {journal} {\bibinfo  {journal} {Phys. Rev. D}\ }\textbf {\bibinfo {volume} {65}},\ \bibinfo {pages} {084014} (\bibinfo {year} {2002})},\ \Eprint {http://arxiv.org/abs/hep-th/0109133} {arXiv:hep-th/0109133} \BibitemShut {NoStop}%
\end{thebibliography}
\end{document}